\documentclass[twocolumn,superscriptaddress,preprintnumbers,amsmath,amssymb,nofootinbib]{revtex4}
\usepackage{graphicx}
\usepackage{textcomp}   
\usepackage{scrextend}   
\usepackage[scaled]{helvet}  
\usepackage{amsmath,amssymb}  
\usepackage{amsfonts}
\usepackage{gensymb}
\usepackage{float}
\usepackage{mathrsfs}
\usepackage{type1cm}
\usepackage{bm}  
\usepackage{color}
\usepackage{slashed}
\usepackage{tikz}
\usetikzlibrary{arrows,calc,decorations.markings,math,arrows.meta}
\usetikzlibrary{shapes}
\tikzset{My Arrow Style/.style={single arrow, fill=orange!50, anchor=base, align=center,text width=2.8cm}}

\usepackage{fixltx2e}
\usepackage{cancel}
\usepackage{bbding}
\usepackage[breaklinks,colorlinks=true]{hyperref}
\usepackage[absolute,overlay]{textpos}
\newcommand{\bea}{\begin{eqnarray}}
\newcommand{\eea}{\end{eqnarray}}

\begin{document}

\title{Symmetry Breaking and Lattice Kirigami}

\author{Eduardo V. Castro} 
\affiliation{CeFEMA, Instituto Superior T\'{e}cnico, Universidade de Lisboa, Av. Rovisco Pais, 1049-001 Lisboa, Portugal}
\affiliation{Centro de F\'isica das Universidades do Minho e Porto, Departamento
de F\'isica e Astronomia, Faculdade de Ci\^{e}ncias, Universidade do Porto,
4169-007 Porto, Portugal}
\affiliation{Beijing Computational Science Research Center, Beijing 100084,
China}
\author{Antonino Flachi} 
\affiliation{Department of Physics \& Research and Education Center for Natural Sciences, Keio University, 4-1-1 Hiyoshi, Kanagawa 223-8521, Japan}
\author{Pedro Ribeiro} 
\affiliation{CeFEMA, Instituto Superior T\'{e}cnico, Universidade de Lisboa, Av. Rovisco Pais, 1049-001 Lisboa, Portugal}
\author{Vincenzo Vitagliano} 
\affiliation{Department of Physics \& Research and Education Center for Natural Sciences, Keio University, 4-1-1 Hiyoshi, Kanagawa 223-8521, Japan}

\date{\today}
\begin{abstract}
In this work we consider an interacting quantum field theory on a curved two-dimensional manifold that we construct by geometrically deforming a flat hexagonal lattice by the insertion of a defect. Depending on how the deformation is done, the resulting geometry acquires a locally non-vanishing curvature that can be either positive or negative. Fields propagating on this background are forced to satisfy boundary conditions modulated by the geometry and that can be assimilated by a non-dynamical gauge field. We present an explicit example where curvature and boundary conditions compete in altering the way symmetry breaking takes place, resulting in a {surprising} behaviour of the order parameter in the vicinity of the defect. The  
effect described here is expected to be generic and of relevance in a variety of situations. 
\end{abstract}

\maketitle
\textit{Introduction.} The theory of quantum fields in curved space has produced over its more than fifty years of existence remarkable results \cite{BD,PT}, the phenomena of particle production in gravitational fields \cite{Parker:1969au} and that of black hole evaporation \cite{Hawking:1974sw} being, without doubt, amongst its most celebrated offsprings. 
From a more general perspective, its semi-classical framework has established a highly non-trivial connection between thermodynamics, gravity, and quantum field theory, and it is at this crossroad where non-trivial manifestations of the geometrical and topological attributes of curved space on the quantum domain occur.

A particularly interesting corner of this intersection is that of quantum field theories featuring spontaneous symmetry breaking, where effects of curved space are expected to alter the way vacuum destabilization and phase transitions take place. In absence of gravity, that is in flat space, the story has been known for a long time and the Coleman-Weinberg mechanism has clarified the way radiative corrections may destabilise the vacuum, with symmetries being spontaneously broken as a result of quantum effects \cite{Coleman:1973jx}. 

On curved backgrounds there are differences that are not difficult to anticipate. A first indication on how things change comes from the same Coleman-Weinberg mechanism that, in flat space, predicts a first-order phase transition from a broken to a restored symmetry phase in scalar electrodynamics, when the scalar mass is increased \cite{Coleman:1973jx}. When lifted to a weakly curved space, renormalization theory implies the appearance of mass-like contribution proportional to the Ricci curvature, thus causing an effective increase of the mass of the scalar. It is then natural to expect that the effect of a positive (negative) spacetime curvature would be to push the system {\it towards} a phase of unbroken (broken) symmetry. 

Additional insight comes from considering spacetimes with horizons that are periodic in Euclidean (imaginary) time. In such a situation, Green's functions enjoy this periodicity with the period set by the horizon size, in analogy with thermal Green's functions that share the same periodicity, but with the period set by the inverse temperature. This leads to the 
expectation that for a sufficiently small horizon a transition from a broken to a symmetric phase may occur. 

These arguments have been made quantitative in a number of cases. Some of the initial discussions focused on scalar fields and spatially homogeneous backgrounds (e.g., de Sitter space) and can be found in Refs.~\cite{Gibbons:1978gg,Shore:1979as,Denardo:1982yi}, where the direct evaluation of the effective potential has shown that a positive curvature does indeed assists symmetry restoration. Interestingly, it was also shown that the details of the scalar field theory (its conformal invariance or lack of it) were responsible for a change in the order of the curvature-induced phase transition from first to second order \cite{Shore:1979as}. Similar issues in relation to chiral symmetry breaking have also been discussed and an extensive review of earlier works is given in Ref.~\cite{Inagaki:1997kz}. The situation becomes more complicated when the background is inhomogeneous or topologically non-trivial. A sample of early calculations in topologically non-trivial spacetimes can be found in Refs.~\cite{Isham:1977yc,Toms:1979ij,Ford:1981xj,Kennedy:1981gu}. In relation to spatially varying geometries a particularly interesting example is that of black holes, for which the question of symmetry restoration was discussed, for instance, in Refs.~\cite{Hawking:1980ng,Moss:1984zf}. There it was shown that a spontaneously broken symmetry is locally restored near a (sufficiently hot) black hole (see also Ref.~\cite{Flachi:2011sx,Flachi:2015fna}). Once again, the interpretation is that the strong gravitational gradient near the horizon is responsible for inducing symmetry restoration.  

The natural playground for contemplating how spacetime topology and curvature might modify the stability of the vacuum in quantum field theory has always been domain of early universe cosmology. Recently, however, other areas of physics are contributing to modernise the above questions and to formulate new exciting problems. 

One such area is related to recent advances in condensed matter research at the nanoscale, particularly in connection with layered materials. Graphene and, more generally, two-dimensional materials are the most spectacular example of the sort, owing to geometrical versatility coupled to an emergent relativistic behaviour of fermions \cite
{NetoRev,VozmRev,AmorimRev}. In these examples, the background geometry is the two-dimensional lattice on top of which fluctuations propagate and the relevance of curvature effects has already been appreciated \cite
{Cortijo:2011aa,Sitenko:2007ewb,Cortijo:2006xp,deJuan:2010zz,Iorio}. Other interesting examples can be found in Refs.~\cite{Grushin,Gromov}.

QCD physics is also fuelling novel research where the use of language and methods of quantum field theory in curved space is becoming more common. Some interesting examples range from the more generic remarks of Refs.~\cite{Flachi:2014jra,Flachi:2015sva} (and references therein), to the very popular area of strongly interacting fermions and chiral symmetry breaking in rotating backgrounds (see, for example,\cite{Chen:2015hfc,Ebihara:2016fwa,Jiang:2016wvv,Chernodub:2017mvp,Flachi:2017vlp,Flachi:2017vlp2,Huang:2017pqe}), to applications of lattice QCD (see, for example, \cite{Yamamoto:2013zwa,Villegas:2014dqa,Yamamoto:2014vda}). In these contexts a range of peculiar geometry-induced phenomena are expected to occur (e.g, condensate suppression/enhancement, appearance of new phases, changes in the critical points geography), whose physical relevance spans from relativistic heavy ion collisions, to transport phenomena, to the astrophysics of compact stars.

The focus of this paper is to reconsider the role of the background geometry in affecting the stability of the vacuum. We will argue that, contrary to expectation, increasing the spatial curvature does not necessarily imply that the system moves closer to a phase of restored symmetry, and we shall present an explicit example of this. Although the example is non-trivial, it is amenable to simple explanation and anticipates the possibility of appearance of exotic changes in the phase behaviour of interacting quantum field theories with a number of interesting implications that we will mention later. 

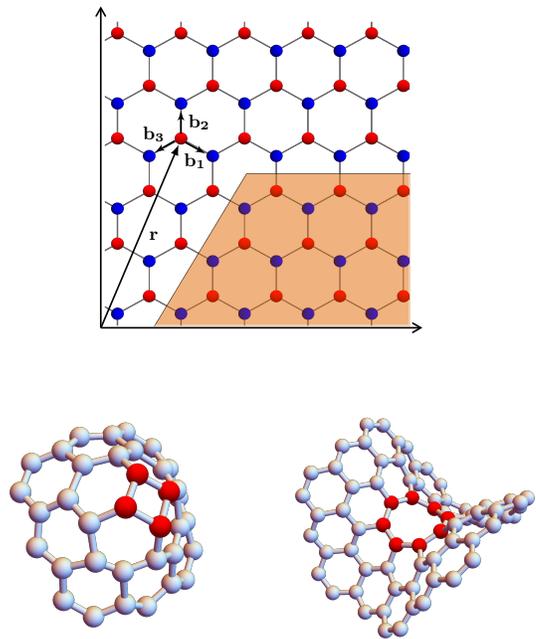
\begin{figure}
\vspace{-10pt}\resizebox{0.3\textwidth}{!}{\begin{minipage}[t]{.8\linewidth}\vspace{0pt}
 \includegraphics[width=\linewidth]{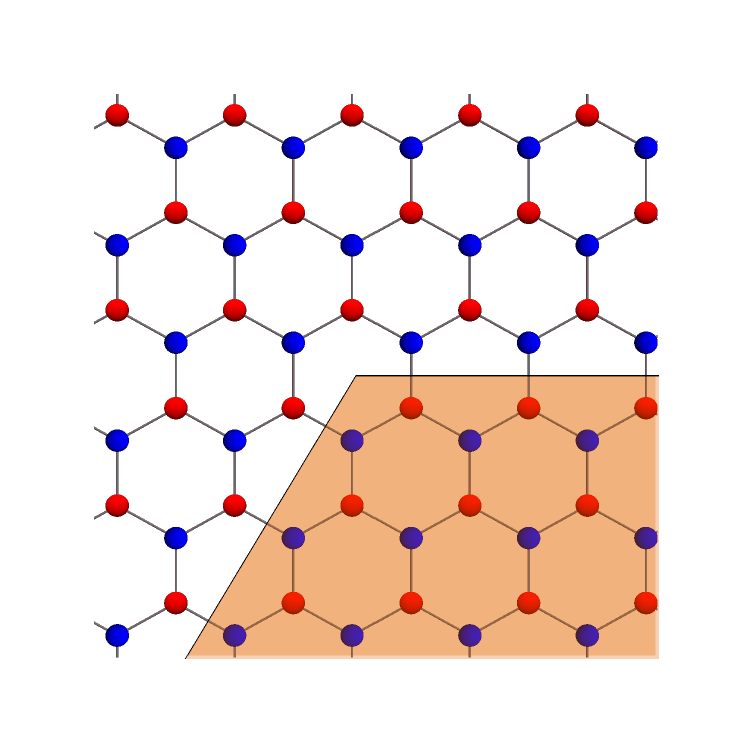}
   \put(-175,23){\begin{tikzpicture}[thick]  
\draw[-{angle 60[scale=3.0]}] (0,0) -- (5.5,0);
\end{tikzpicture}}
   \put(-175,23){\rotatebox{90}{\begin{tikzpicture}[thick]  
\draw[-{angle 60[scale=3.0]}] (0,0) -- (5.5,0);
\end{tikzpicture}}}
\put(-134,113.5){\rotatebox{-29}{\begin{tikzpicture}[thick]  
\draw[-{latex'[scale=3.0]}] (0,0) -- (0.4,0);
\end{tikzpicture}}}
\put(-134,102){\bf{{b\textsubscript{1}}}}
  \put(-149,115){\rotatebox{210}{\begin{tikzpicture}[thick]  
\draw[-{latex'[scale=3.0]}] (0,0) -- (0.4,0);
\end{tikzpicture}}}
\put(-132,121){\bf{{b\textsubscript{2}}}}
 \put(-136,117.5){\rotatebox{90}{\begin{tikzpicture}[thick]  
\draw[-{latex'[scale=3.0]}] (0,0) -- (0.4,0);
\end{tikzpicture}}}
\put(-154,115){\bf{{b\textsubscript{3}}}}
 \put(-175,23){\rotatebox{67}{\begin{tikzpicture}[thick] 
\draw[-{latex[scale=3.0]}] (0,0) -- (3.4,0);
\end{tikzpicture}}}
\put(-151,66){\bf{{r}}}
 \end{minipage}}
\begin{minipage}[t]{.45\linewidth}
   \includegraphics[width=\linewidth]{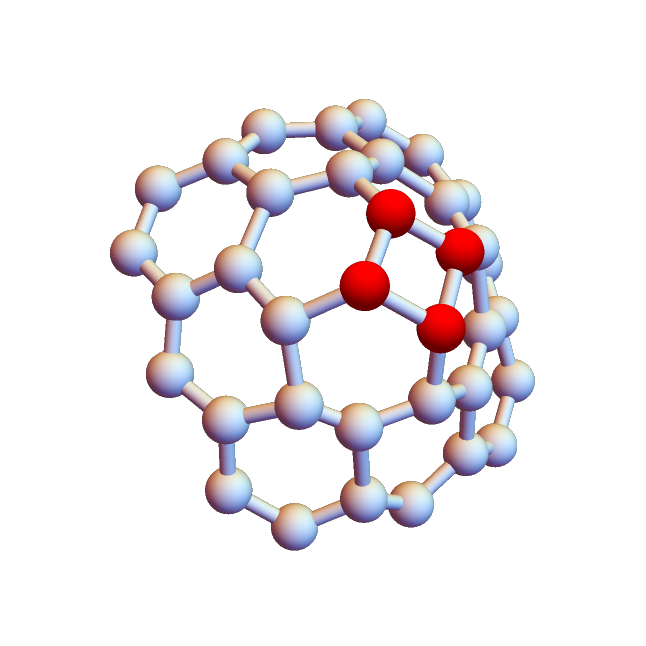}
 \end{minipage}
 \begin{minipage}[t]{.45\linewidth}
\includegraphics[width=\linewidth]{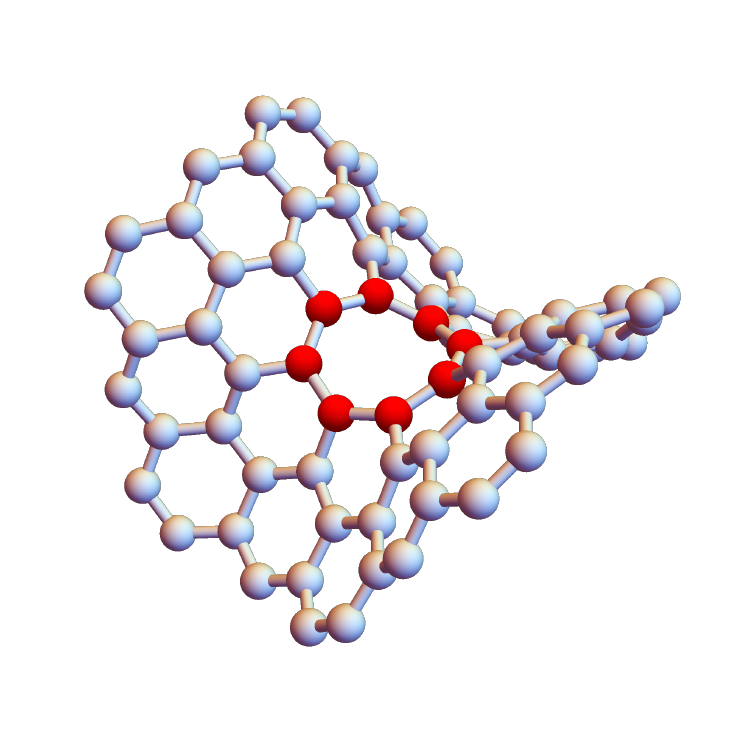}
 \end{minipage}
 \caption{
A honeycomb, planar lattice with bipartite nature, and some possible defective configurations. The top figures illustrate the flat hexagonal lattice with a $2\pi/3$ section highlighted. The exchange symmetry between the two triangular (red and blue) sub-lattices is also illustrated. Defects are introduced by a procedure of adding or cutting the lattice $\pi/3$ sections and gluing the remaining sides along the cut. For example, subtracting a $2\pi/3$ section, one obtains a positive curved cone with a $n_s=4$-sides defect (bottom left). Adding a $2\pi/3$ section, instead, generates a negative curved saddle geometry with $n_s=8$ (bottom right). These lattice structures are well known and form the basis of chiral curved poly-aromatic systems (n-circulene, see Ref.~\cite{circulene} for a review of the geometry of these lattice structures).}
   \label{Figura1}
\end{figure}

\textit{Model and geometry.} For the sake of concreteness, we shall consider here a specific class of $(2+1)$-dimensional interacting field theoretical models of the Hubbard-type, whose Hamiltonian ${\mathsf H} = {\mathsf H}_0+{\mathsf H}_I$ is expressed as the sum of a free part,
\begin{equation}
{\mathsf H}_0 = - t \sum_{{\bf r},~{i},~\sigma=\pm} u^\dagger_\sigma({\bf r}) v_\sigma({\bf r}+{\bf b}_i) + \mbox{H.C.},\nonumber
\end{equation}
plus an interacting sector,
\begin{equation}
{\mathsf H}_I = 
{U\over 4} \sum_{{\bf r},\sigma,\sigma',i} \left( 
n_\sigma({\bf r})n_{\sigma'}({\bf r}) 
+
n_\sigma({\bf r}+{\bf b}_i)
n_{\sigma'}({\bf r}+{\bf b}_i)
\right).\nonumber
\end{equation}
The above field theory is defined on an underlying lattice that we assume for the moment to be flat with hexagonal cells and generated by linear combinations of a set of basis vectors as illustrated in Fig.~\ref{Figura1} (${\bf r}$ span a triangular sub-lattice and the vectors ${\bf b}_i, i=1,2,3$ connect the atom in ${\bf r}$ with the three nearest-neighbours). The annihilation operators of the two sub-lattices are $u$ and $v$ and $n_\sigma$
is the number operator. The quantities $t$ and $U$ are positive numbers describing, respectively, the hopping and the interaction constant.

The above model is routinely used to describe many of the properties of graphene and other layered materials \cite{NetoRev,KotovRev}. An important aspect is the possibility to locally induce curvature by deforming the lattice with the insertion of defects. Also, the continuum limit is not difficult to analyse and generalisations can be easily imagined. Finally, although here we will be concerned with the continuum limit, carrying out lattice simulations should be feasible. 

The specific type of symmetry breaking that we wish to discuss here is associated with the bipartite nature of the honeycomb lattice that the Hubbard model above captures in the magnetisation that we shall properly define below. Since our goal here is to scrutinise the effect of curvature on the spontaneous breakdown of the above sub-lattice symmetry, our first task is to covariantize the model to curved space. For this it is convenient to work with the continuum Lagrangian counterpart that can be obtained using standard methods by expressing the original Hamiltonian in terms of the SU$(2)$ vector ${\bf S} =\sum_{\sigma,~\sigma'} u^\dagger_\sigma({\bf r}) \vec{\tau}_{\sigma,\sigma'} u_{\sigma'}({\bf r})/2$, where $\vec{\tau}$ is a vector with the Pauli matrices as components, and then proceed by means of a Hubbard-Stratonovich transformation. In order to maintain our treatment as simple as possible we will assume a scalar order parameter that can be motivated by a {rotational anisotropy} favouring symmetry breaking along the $z$ axes (for graphene this could be due to the presence of a substrate and related spin-orbit coupling enhancement \cite{rossier,morpurgo}). This allows to gap out the Goldstone modes that can be straightforwardly included in a more involved treatment. Choosing an auxiliary field $\phi$ that breaks both the $\mathbb{Z}_2$ and the discrete sub-lattice symmetry, the Hamiltonian ${\mathsf H}$ can be mapped, at low energies, onto the following $(2+1)$-dimensional field theory
\begin{equation}
{\mathscr L} = \bar{\psi}_\sigma \imath \slashed{\partial} \psi_\sigma + \left(\sigma \bar{\psi}_\sigma \phi \psi_\sigma\right) +{\phi^2\over 2\lambda}\;,
\label{lag}
\end{equation}
where the first term is a free Dirac contribution and the remaining terms describe the interaction sector. 
The summation over repeated spin indices $\sigma = \pm$ is understood and the four-component spinors $\psi_\sigma$ are arranged as $\psi_\sigma^T = \left(\psi^{A1}_{\sigma}, \psi^{B1}_{\sigma}, \psi^{A2}_{\sigma}, \psi^{B2}_{\sigma} \right)$, with $\psi^{IJ}_\sigma(x) = {a \over v_F} \int {d^2p \over (2\pi)^2} \; e^{-\imath {\bf p} \cdot {\bf x}} z^{IJ}_\sigma ({\bf p})$ and where $z^{I,J}_\sigma(p) = z^I_\sigma({\bf K}_J + {a \over v_F} {\bf p} ) $ represents the sub-lattice annihilation operators ($z^{A}=u, z^{B}=v$) near the two Dirac cones  ${\bf K}_{J =1, 2}$ of the 
dispersion relation. The spatial coordinates were rescaled by ${\bf  x} =  {{\bf r} / v_F} $ where $v_F=3/2 t a$ is the Fermi velocity and $a$ is the lattice spacing. 
Finally, the coupling constant $\lambda$ is proportional to the interaction strength $\lambda \propto U$ up to an unimportant factor,  dependent on the particular regularization of the low energy theory.

The exchange of the sub-lattices can then be implemented by the 
simultaneous exchange $x_2 \rightarrow -x_2$ ($p_2 \rightarrow p_2$), leaving intact the Dirac points and the spin, and the Lagrangian (\ref{lag}) invariant as long as $\phi$ vanishes. The order parameter for the above symmetry is ${\phi} = 2\lambda \langle \bar\psi_- \psi_- - \bar\psi_+\psi_+ \rangle$ and it describes the staggered magnetization, i.e., $\phi \neq 0$ indicates broken symmetry. It is possible to arrive at the same expression (\ref{lag}) following the general decomposition of the Hubbard Hamiltonian as outlined in \cite{Herbut}.

By means of a kirigami like procedure\footnote{Kirigami is a variation of origami that includes cutting of the paper, rather than solely folding the paper \cite{wikipedia}.} we introduce a spatial curvature in the model by inserting a disclination that warps the lattice locally. There are many ways to do this, but if we wish to isolate the interplay between quantum effects and geometry, we need to preserve the bipartite nature of the lattice at tree level, that is avoid frustrating the lattice. This requirement restricts the allowed deformations to those induced by defects with an even number of sides, as these are the only that preserve the above symmetry classically. Inserting a defect with $n_s<6$ sides in an hexagonal lattice generates a deficit angle and a curvature that is locally positive (see Fig.~\ref{Figura1}). In contrast, adding a defect with $n_s>6$ generates an excess angle and a locally negative curvature (see Fig.~\ref{Figura1}). 

In the continuum model, the curvature can be introduced by specifying the background metric to be that of a manifold with a conical singularity. The Riemannian geometry of such manifolds is studied since, at least, \cite{Sommerfeld}. Refs.~\cite{Fursaev:1995ef,Fursaev:2001og} give details and additional bibliography on the topic. Here, in order to model such a localised curvature, we use a Euclidean parametrisation for the metric tensor
\begin{equation}
ds^2 = d\tau^2 + dr^2 +  \alpha^2 r^2 d\theta^2
\label{cone}
\end{equation}
with $r \geq 0$ and $0\leq \theta < 2\pi$ being the polar coordinates centred at the apex. We will not concern ourselves with finite temperature effects here, but these can be included in a straightforward manner by using the standard imaginary time formalism. 
Defining $\tilde\theta = \alpha\; \theta$, it should be clear that the metric is that of flat space with $0\leq \tilde\theta < 2\pi \alpha$. If $\alpha<1$, then $\gamma = 2\pi -  2\pi \alpha$ describes a deficit angle. Removing the deficit angle and identifying the two sides results in a cone with opening angle $2 \arcsin \alpha$. The closer to unity is $\alpha$, the flatter is the cone. If $\alpha >1$, then the deficit angle becomes an excess angle.

Since the curvature of conical manifolds diverges at the apex, some regularisation is necessary to deal with the singular behaviour. Here, we will regulate the geometry by replacing the singular space with a sequence of regular manifolds as done in Ref.~\cite{Fursaev:1995ef,Fursaev:2001og}. Calculations are done in the regularised geometry and results in the original singular space are obtained as a limit, once the regularisation is removed at the end. 
In practise, this procedure can be implemented by replacing the original metric (\ref{cone}) with the following regular one:
\begin{equation}
d\tilde{s}^2 = d\tau^2 + 
f_\epsilon(r) dr^2 + \alpha^2 r^2 d\theta^2
\label{dsa2}
\end{equation}
where $\epsilon$ represents a regularisation parameter and $f_\epsilon(r)$
is a smooth function satisfying the following properties: 
1) $\lim_{\epsilon \rightarrow 0} f_\epsilon(r) = 1$; 
2) $f_\epsilon(r) \approx 1$ for $r \gg \epsilon$; 
3) $f_\epsilon(r) = \mbox{const}$ for $r = 0$. 
It should be noted that while the limit of $\epsilon \rightarrow 0$ corresponds to removing the regularisation, 
in an eventual comparison with a lattice simulation, $\epsilon$ should be associated with the lattice spacing and it acquires the status of a physical cut-off. 

The Lagrangian (\ref{lag}) 
is extended to curved space by a minimal covariantization procedure, i.e. letting the Minkowski metric, the derivatives, and the gamma matrices to the corresponding quantities in curved space. We shall not include non-minimal couplings in the present treatment (see Ref.~\cite{Flachi:2010yz} for a discussion about this point). 

The last element we need to take into account are the boundary conditions along the cut where the two sides of the lattice have been glued after having removed or added a portion of the lattice to accomodate the insertion of the defect. The same procedure that we shall use below has been discussed, for example, in Ref.~\cite{Sitenko:2007ewb}. It is not difficult to realise that for a generic even-sided defect, the two sub-lattices are unchanged and the fermion wave function, after circulating around the defect, satisfies the following boundary condition: $\psi(r,\varphi+2\pi) = - \exp{\left(i {(6-n_s)\pi \gamma_5/2} \right)}\psi(r,\varphi)$.
(Here, we follow the same conventions as Ref.~\cite{Sitenko:2007ewb}, and choose to work in the standard planar representation of the Clifford algebra of $\gamma$-matrices, where $\gamma_0$ is diagonal). A transparent way to incorporate these boundary conditions is by re-expressing the fields as
$\psi'(r,\varphi) = \exp{\left(-i \varphi {(6-n_s) \gamma_5/4} \right)}\psi(r, \varphi)$, and by noticing that the primed fields obey the standard periodicity condition $\psi'(r,\varphi+2\pi) = - \psi'(r,\varphi)$. It is simple to prove that the effect of the above redefinition is to augment the Lagrangian by a non-dynamical gauge connection $\mathscr{A}_\mu = - \delta_\mu^\varphi (6-n_s)\gamma_5/4$. This term will be crucial in altering the way symmetry breaking takes place.

{\it Methods and results.} With all of the above in hands, we can examine whether the geometry and associated boundary conditions, which we have implemented as described in the preceding section, favour a phase of broken or restored symmetry in the region where curvature attains a positive (in the case of $n_s=4$ and deficit angle) or negative (in the case of defects with even $n_s>6$ and excess angle) value.
As we have motivated at the beginning, since the curvature increases (decreases) as we approach the defect for $n_s=4$ ($n_s=8,10,\dots$), the expectation is that curvature should favour symmetry restoration in the vicinity of the defect for $n_s=4$, while for $n_s$ even and larger than $6$ broken symmetry should instead be favoured. 

Below we shall address this question by computing the effective action for the order parameter $\phi$ and by numerically solving the associated effective equations. There are a few technical steps that we should clarify in order to allow anyone to reproduce our results. First of all, our analysis follows the large-$N$ deformation of Ref.~\cite{Herbut}, where we pass from 2 to $N$ flavours of the Dirac fields: $\bar\psi_+\psi_+ \rightarrow \sum_{\sigma=1}^{N/2}\bar\psi_\sigma\psi_\sigma$ and $\bar\psi_-\psi_- \rightarrow \sum_{\sigma=N/2+1}^{N}\bar\psi_\sigma\psi_\sigma$. Then, the effective action 
at mean-field level can be expressed as 
\begin{equation}
\tilde{\Gamma} \left[\phi \right] = - \int d^3x \sqrt{\tilde{g}} \frac{\phi^2}{2\lambda} + {1\over 2} \sum_{p=\pm} \log \det \left( \tilde{\Box} + {\tilde{R}\over 4} + {\phi}_p^2\right),
\nonumber
\end{equation}
where ${\phi}_\pm^2=\phi^2 \pm \sqrt{\tilde{g}^{rr}}\phi'$ and the D'Alembertian is calculated from the spinor covariant derivative $\tilde D_\nu = \tilde\nabla_\nu + i \mathscr{A}_\nu$. The tildes indicate that quantities are computed using the regularised metric (\ref{dsa2}). 

Heat-kernel methods and zeta regularization techniques are used to perform the computation of the determinant above. The interested reader is referred to Refs.~\cite{PT,BOS,elizalde,avramidi,kirsten}) for general background and to Refs.~\cite{Flachi:2010yz,Flachi:2011sx} where similar calculations have been done in the context of interacting fermion effective field theories in curved space. The algebra is handled by computer symbolic manipulation, here we summarise the essential steps: 1) we first express the determinant in terms of a covariant derivative expansion of the propagator in powers of curvature invariants and derivatives of $\phi$ (with scalar curvature contributions fully resummed as in Refs.~\cite{Parker:1984dj,Flachi:2010yz}); 2) we truncate this expansion to second order in the heat-kernel expansion and write the effective equations for $\phi$ arising from this truncated effective action; 3) finally, we use numerical approximation to solve the resulting effective equations.   

\begin{figure}
\begin{center}
\includegraphics[width=0.44\textwidth]{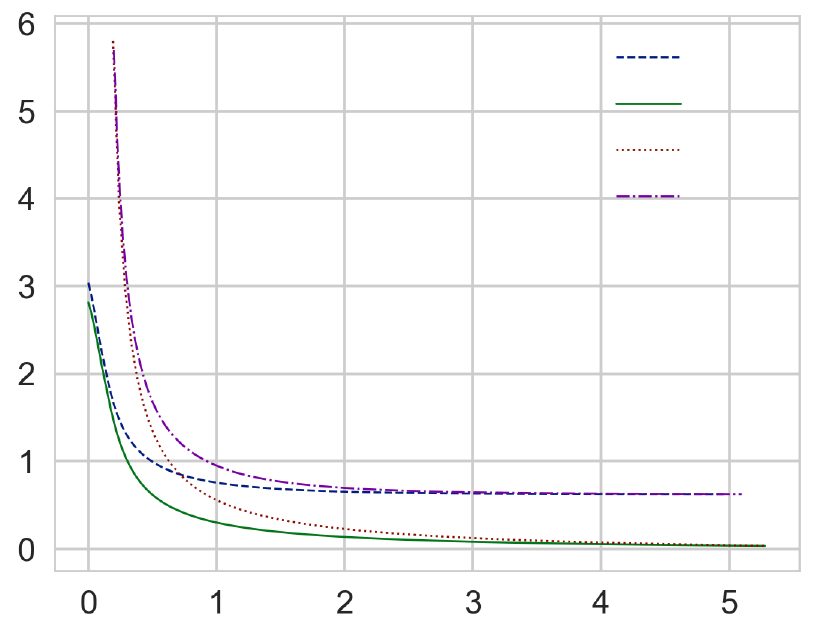}
\put(-37,152){\tiny{$\lambda_s=10$}}
\put(-37,140){\tiny{$\lambda_o=500$}}
\put(-37,128){\tiny{$\lambda_s=500$}}
\put(-37,116){\tiny{$\lambda_o=10$}}
\put(-240,90){$\phi$}
\put(-110,-10){$r$}
\end{center}
\caption{The curves are obtained by numerical approximation and illustrate how the order parameter $\phi$ changes as the defect is approached. We have set the values of the coupling constant $\lambda_{s,o}$, in units of the renormalisation scale, to encompass both the cases in which far away from the defect the system is in a phase of broken or unbroken symmetry. The indices $s$ and $o$ stand for square and octagon. In the numerical computation we have varied the regularization parameter $\epsilon$ from $0.1$ down to $0.01$.}
\label{Figura2}
\end{figure}
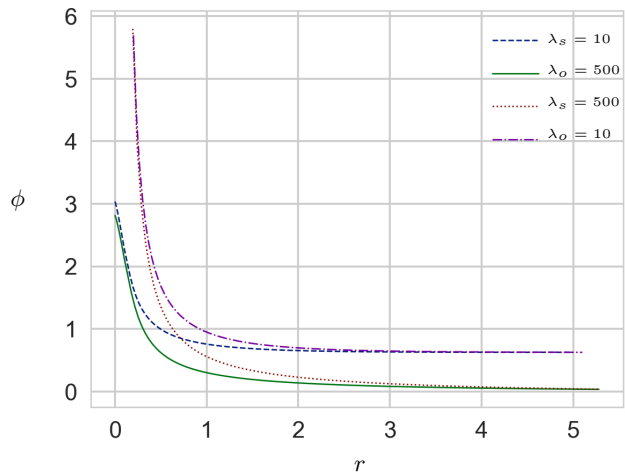

The results are illustrated in Fig.~\ref{Figura2} for a few cases with both locally positive ($n_s=4$) and negative curvature ($n_s=8$). The asymptotic value of the coupling constant 
should be fixed by imposing specific renormalization conditions (see Ref.~\cite{Inagaki:1997kz}), with its value adjusted to specific situations. In the present calculation, we have changed its value in order to encompass both situations in which symmetry is either broken or close to the critical value far away from the defect (see Fig.~\ref{Figura2}). The numerical solutions show that $\phi$ develops a spatial variation and attains a value near the defect that is larger than its asymptotic value for any $n_s \neq 6$, signalling that in the vicinity of the defect curvature, irrespectively of its sign, seems to encourage an ordered phase.

To examine what is going on, let's consider for illustration the case of $n_s=4$, corresponding to a locally positive scalar curvature. In such a case, intuition suggests that a positive curvature ($\tilde R > 0$) should drive the system towards a symmetric phase with the order parameter moving towards smaller values, thus making the observed behaviour rather unexpected. However, as we have seen earlier, the boundary conditions along the cut also play an essential role by inducing the emergence of a non-dynamical gauge field $\mathscr{A}_\mu$, whose origin is entirely geometrical (the number of sides of the defect can be interpreted as an effective charge) and whose effect, in the fermion determinant, combines with that of the scalar curvature. While the term proportional to $\tilde R$ is responsible for pushing the order parameter towards a symmetric phase in conformity with the arguments of Ref.~\cite{Flachi:2014jra}, the shift caused by $\mathscr{A}_\mu$ competes with the effect of $\tilde R$. In addition, such a term appear with a modulation factor that originates from the metric tensor, $\sim g^{\mu\nu}\mathscr{A}_\mu\mathscr{A}_\nu \propto - n_s^2/r^2$, that amplifies its effect near the defect, explaining the observed behaviour.

{\it Conclusions.} In this work we have engineered a curved background starting from a flat 2D hexagonal lattice using a kirigami-like procedure of removing (adding) a piece of lattice and by gluing the parts along the cut, a way to geometrically inserting a defect in the lattice. In the continuum this curved lattice corresponds to a Riemannian manifold with a locally positive (negative) curvature and a conical-like singularity at the defect. We have considered an interacting quantum field theory on this background and analysed how symmetry breaking is altered by the geometrical deformation caused by the defect. 

The two important features of the story turn out to be the increasing (or decreasing) curvature near the defect and the boundary conditions along the cut that can be assimilated by a non-dynamical gauge field modulated by the conical structure. As a working example, we have looked at the staggered magnetisation, the order parameter associated with the discrete sub-lattice symmetry. The numerical results have shown an increase of the order parameter as the locally curved region is approached, a behaviour that signals a change towards order and that goes against the expectation that increasing the curvature should drive the system closer to a state of unbroken symmetry. This behaviour has been explained by the competition between curvature and the emergent gauge field, i.e. between geometry and a feature of topological nature related to the boundary conditions. 

The kirigami effect we have described should be generic and naturally expected to occur for different lattice structures (i.e., different unit cells) as long as the same geometrical traits are maintained. We also expect the same to occur for different field theory models. One intriguing possibility is to consider a multi-defect configuration and see whether there is any special arrangement where the relative weight of the geometry-induced gauge fields {\it vs} curvature can be adjusted, thus tailoring specific configurations of the order parameter.

{Amongst the various interesting implications, the positive effect of curvature on the spontaneous breaking of sub-lattice symmetry could be used to shed light on the long lasting question regarding the semimetal-insulator phase transition in graphene: even though graphene is predicted to  be very close to the transition point \cite{KatsSIPT, KotovRev}, no experimental signature of the insulating behaviour has been found in flat graphene so far \cite{GeimCloseDir}. Another promising route in graphene would be the combination of curvature with adatom adsorption in order to enhance symmetry breaking, in particular of magnetic order \cite{PalaciosH}.}

An interesting direction to extend the idea of this work is to look at higher dimensionality. A straightforward application is to (straight) cosmic strings 
\cite{VS}. In this case, the effect discussed here should trigger fermion condensation at the string. This would offer a mechanism of inducing a superconducting phase of different nature from usual arguments (see \cite{VS}). The idea should deserve some attention both in cosmology and condensed matter physics (e.g., liquid crystals; see, for example, \cite{Nelson,Vitelli}).

While beyond of the scope of this work, it is tempting to relate the present ideas to gravity at the Planck scale, where spacetime may be discrete. In this case, the presence of defects in the background lattice would cause local changes in the geometry and topology, similar to those described here, that could trigger a form of graviton condensation in the vicinity of these spacetime glitches. 

\textit{Acknowledgments.} The support of the Japanese Ministry of Education, Culture, Sports, Science Program for the Strategic Research Foundation at Private Universities `Topological Science'  (Grant No.\ S1511006) is gratefully acknowledged. VV is supported by the Japanese Society for Promotion of Science (Grant No. P17763). EVC and PR acknowledge partial support from FCT-Portugal through Grant No. UID/CTM/04540/2013. AF is grateful to A. Beekman, K. Fukushima, T. Fujimori, M. Nitta, R. Yoshii for discussions on various aspects of symmetry breaking and geometry and to A. Beekman for carefully reading the manuscript and for various suggestions.

\end{document}